\documentclass[aps,prb,preprint,superscriptaddress]{revtex4-1}
\usepackage{color}
\usepackage{textcomp}
\usepackage{graphicx}
\usepackage{lineno}
\usepackage[normalem]{ulem}
\usepackage[colorinlistoftodos]{todonotes}
\begin{document}
\title{Tuning spin excitations in magnetic films by confinement}

\author{Jonathan Pelliciari}
\email{pelliciari@bnl.gov}
\affiliation{National Synchrotron Light Source II, Brookhaven National Laboratory, Upton, NY 11973, USA.}

\author{Sangjae Lee}
\affiliation{Department of Physics, Yale University, New Haven, Connecticut 06520, USA.}

\author{Keith Gilmore}
\affiliation{Condensed Matter Physics and Materials Science Department, Brookhaven National Laboratory, Upton, New York 11973, USA.}

\author{Jiemin Li}
\affiliation{National Synchrotron Light Source II, Brookhaven National Laboratory, Upton, NY 11973, USA.}

\author{Yanhong Gu}
\affiliation{National Synchrotron Light Source II, Brookhaven National Laboratory, Upton, NY 11973, USA.}

\author{Andi Barbour}
\affiliation{National Synchrotron Light Source II, Brookhaven National Laboratory, Upton, NY 11973, USA.}

\author{Ignace Jarrige}
\affiliation{National Synchrotron Light Source II, Brookhaven National Laboratory, Upton, NY 11973, USA.}

\author{Charles H. Ahn}
\affiliation{Department of Applied Physics, Yale University, New Haven, Connecticut 06520, USA.}

\author{Frederick J. Walker}
\affiliation{Department of Applied Physics, Yale University, New Haven, Connecticut 06520, USA.}

\author{Valentina Bisogni}
\email[]{bisogni@bnl.gov}
\affiliation{National Synchrotron Light Source II, Brookhaven National Laboratory, Upton, NY 11973, USA.}


\begin{abstract}
\textbf{Spin excitations of magnetic thin films are the founding element for novel transport concepts in spintronics, magnonics, and magnetic devices in general. While spin dynamics have been extensively studied in bulk materials, their behaviour in mesoscopic films is less known due to experimental limitations. Here, we employ Resonant Inelastic X-Ray Scattering to investigate the spin excitation spectrum in mesoscopic Fe films, from bulk-like down to 3 unit cells thick. In bulk-like samples, we find isotropic, dispersive ferromagnons consistent with the dispersion observed by neutron scattering in bulk single crystals. As the thickness is reduced, these ferromagnons survive and evolve anisotropically: renormalising to lower energies along the out-of-plane direction while retaining their dispersion in the in-plane direction. This thickness dependence is captured by simple Heisenberg model calculations accounting for the confinement in the out-of-plane direction through the loss of Fe bonds. Our findings highlight the effects of mesoscopic scaling on spin dynamics and identify thickness as a knob for fine-tuning and controlling magnetic properties in films. }
\end{abstract}
\maketitle

The never-ending demand for energy efficient devices, as well as recent advances in material deposition and nanoscale fabrication techniques, have set the basis for novel branches of electronics \cite{han_spin_2020,han_mutual_2019,cai_ultrafast_2020,manchon_current-induced_2019,wang_magnetization_2019,sun_understanding_2019}. One of these is magnonics in which spin-waves replace charge as the means of signal transport with significant technological advantages: the most crucial being the reduction of heat dissipation. However, establishing spin wave-based transport as a viable solution for the next generation of electronics poses several challenges that need to be resolved to enable logic operations \cite{lenk_building_2011} (\textit{e.g.}  effective spin wave signal generation, propagation, control and detection) \cite{rana_towards_2019, wang_magnetization_2019, han_quantum_2018, wang_spin-wave_2016, bracher_phase--intensity_2016,demidov_control_2009,sun_understanding_2019}. At a fundamental level, there are still many outstanding questions concerning how magnetic properties and excitations evolve and emerge in low dimensional systems, such as thin films \cite{soumyanarayanan_emergent_2016, martin_thin-film_2016, hwang_emergent_2012}. The interplay between confinement, dimensionality, interface/surface hybridization, strain, and exchange-bias, can provide effective tuning of magnetism in such thin films, ultimately leading to exquisite control of spin dynamics. Understanding how such spin dynamics unfold in low-dimensional magnetic systems is thus an essential step towards gaining control of spin wave-based transport phenomena.

From an experimental standpoint, the investigation of how spin dynamics evolve from a bulk-like system down to the nanoscale – the so called mesoscale range, where at least one dimension \textit{d} is comparable to the electron mean free path \cite{gall_electron_2016} - has been limited so far by the lack of suitable probes.
While inelastic neutron scattering (INS) is an established technique to study spin dynamics in bulk magnetic materials  \cite{shirane_spin_1968,mook_spinwave_1969,mook_neutron_1973,perring_highenergy_1991, tranquada_superconductivity_2014}, the low cross-section of neutrons makes difficult its application to meso-scaled systems. Alternatively, spin polarized-electron energy loss spectroscopy (SP-EELS) has been employed in the past to detect spin excitations in few monolayer thick films of 3d-metals \cite{zakeri_asymmetric_2010,zakeri_magnon_2011,zakeri_surface_2013,chuang_impact_2012} but provides insights mostly on the two-dimensional spin dynamics confined to the surface (due to the small electron escape depth), failing to access the properties of the whole film. Other techniques such as Brillouin Light Scattering and Ferromagnetic Resonance spectroscopy have been employed to study spin excitations but are limited to \textbf{q}$\sim0$ \cite{grunberg_brillouin_1982,han_mutual_2019,chumak_magnon_2015}.
Recently, resonant inelastic x-ray scattering (RIXS) has emerged as an alternative tool for the detection of spin excitations and their dispersion in magnetic systems \cite{zhou_persistent_2013,pelliciari_reciprocity_2019,le_tacon_intense_2011,lee_asymmetry_2014,peng_influence_2017,suzuki_spin_2019}. Owing to the large cross section and RIXS resonant character, elementary excitations in meso-scaled materials can be efficiently studied \cite{dean_spin_2012, bisogni_ground-state_2016}. 

Here, we use RIXS to investigate the spin dynamics in thin films of ferromagnetic Fe grown using molecular beam epitaxy (MBE). By mapping out the evolution of the spin excitations over almost two orders of magnitude of thickness across the mesoscale range, we establish confinement as an effective variable to manipulate the spin dynamics. For bulk-like Fe films, RIXS is used for the first time to detect ferromagnons and their isotropic three-dimensional dispersion relation. By progressively reducing the film thickness, we observe spin excitations surviving down to the ultra-thin limit. Furthermore, we uncover the evolution of the spin dispersion as it deviates from bulk behaviour and develops anisotropically between the in-plane and out-of-plane directions: renormalising to lower energies along the [0,0,L] out-of-plane direction while being unchanged along the [H,0,0$^*$] in-plane direction. This anisotropic modification is captured by an isotropic Heisenberg-like model which predicts that the out-of-plane dispersion changes when the film thickness approaches the length scale of the Fe magnetic interaction. Our work therefore sets the basis for selectively gaining control of the spin excitation energy through confinement, enabling the design of optimized materials for future applications. 

\begin{figure}[h!]
\centering
\includegraphics[scale=1.0,clip]{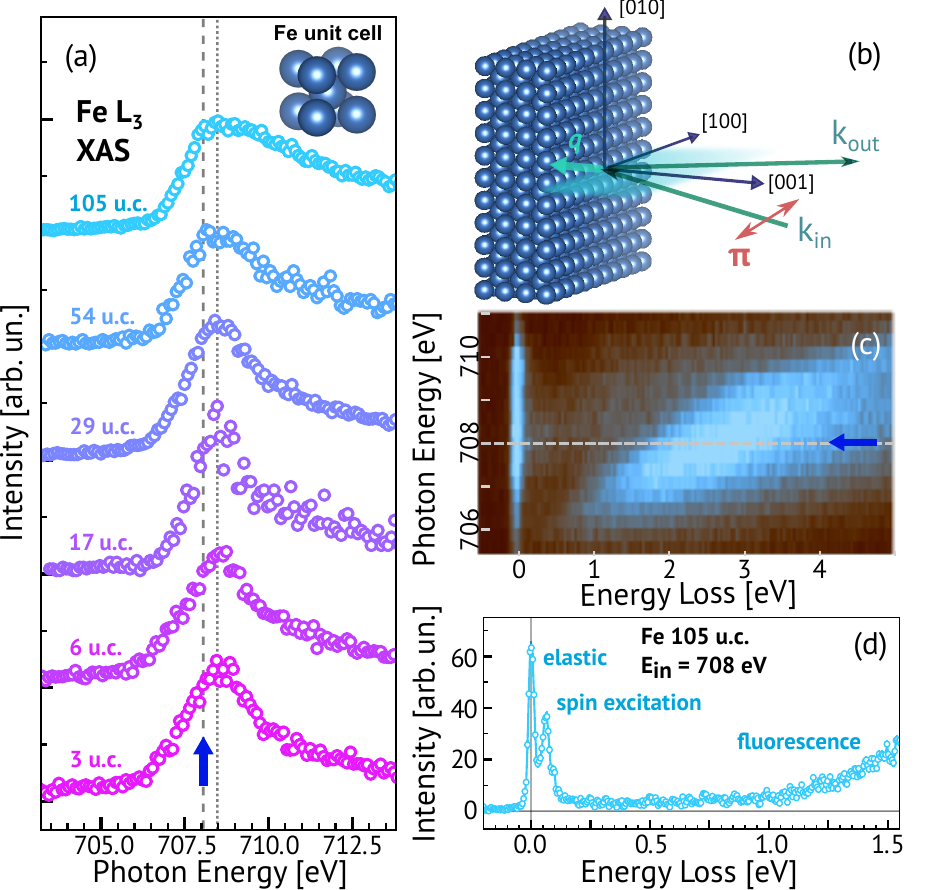}
\caption{\label{fig:fig1} \textbf{Fe-$L_3$ XAS of Fe films versus thickness and representative RIXS measurement from Fe bulk-like film.} (a) Thickness dependence of the Fe-$L_3$ XAS measured in PFY mode. The blue-arrow marks the incident photon energy used for the RIXS spectra.  This energy is defined at -0.4 eV w.r.t. the Fe-$L_3$ $E_{max}$. Top right corner: sketch of the \textit{bcc} unit cell of Fe \cite{momma_vesta_2011}. (b) Experimental configuration used for the RIXS measurements. (c) Incident energy dependence of the RIXS signal across the Fe-$L_3$ edge, for Fe 105~u.c. representative of a bulk-like sample. (d) RIXS spectrum at E$_{in} = 708.0$~eV and (0,0,0.23) for Fe 105 u.c.. }
\end{figure}

High-quality, crystalline films of Fe with varying thickness were prepared by molecular beam epitaxy (MBE). The Fe [001]-oriented films were grown epitaxially on MgO[001] with a 45\textdegree~ rotation around the [001] axis at a substrate temperature of 300 \textdegree{C}.
The thickness of the films, extracted from x-ray reflectivity (see Supplementary Note 1 B), was 105 unit cells (u.c.), 54 u.c., 29 u.c., 17 u.c., 6 u.c., and 3 u.c. (30 to 0.87 nm, respectively). All films retain a body centered structure. The Fe saturation magnetic moment measured by SQUID magnetometry is $\sim2\mu_B$ of all the films (see Supplementary Note 1 C).
Furthermore, considering the relatively large lattice mismatch $\sim3.8\%$ for bulk $\alpha$-Fe (body centered cubic, \textit{bcc}, with lattice parameter a$=2.87$ \AA) grown on top of bulk MgO (diagonal axis 2.98 \AA), we observe strain relaxation as we increase the thickness of the Fe films. In particular, the 29 u.c. film is fully relaxed, while the 3 u.c. film is fully strained to the substrate. This relaxation is evident in the evolution of the measured in-plane lattice parameters: a$_{29uc}=2.89\pm0.01$ \AA, a$_{6uc}=2.97\pm0.01$ \AA, and a$_{3uc}=2.98\pm0.01$ \AA~(refer to Supplementary Note 1 D for the strain study). 

The oxidation state of the films is measured using x-ray absorption spectroscopy (XAS). To achieve XAS with high sensitivity we measured partial fluorescence yield (PFY) using the RIXS spectrometer. The spectra  in Fig.~\ref{fig:fig1}(a) display a broad main peak around $E_{max}\simeq$ 708.4 eV for all films, due to the transition from localized $2p_{3/2}$ core states to the delocalized $3d$ valence states, consistent with metallic Fe \cite{regan_chemical_2001, jimenez-villacorta_x-ray_2011}. The absence of additional peaks, characteristic of iron oxides (FeO, Fe$_2$O$_3$, and Fe$_3$O$_4$), demonstrates a negligible oxidation of Fe.

High-resolution RIXS experiments were performed at the SIX 2-ID beamline of NSLS-II with an energy resolution of $\Delta E = 23$~meV (FWHM) at the Fe-L$_3$ edge \cite{dvorak_towards_2016}. $\pi$-polarized light was used to minimise the elastic scattering signal. All the data were measured at 100~K.
The films were aligned with both the surface normal [001] and the [100] axis lying in the scattering plane as depicted in Fig.~\ref{fig:fig1}(b). The spectra are normalised to the integrated fluorescence to have comparable intensities for all films.

We first show that RIXS probes the spin excitations in bulk-like Fe. Up to now, this technique has never been utilized to study ferromagnetic excitations in metals. We initially collected an incident energy dependence of the RIXS signal across the Fe $L_3$ edge on a bulk-like film. Figure~\ref{fig:fig1}(c) displays the RIXS map for the 105 u.c. film: the most apparent features of the spectra consist of a broad fluorescence line due to the incoherent $3d \rightarrow 2p$ emission and a low-energy ($<100$ meV) Raman-like peak close to the elastic line. We achieve the best contrast between this Raman-like peak and the background by selecting an incident photon energy de-tuned by -0.4 eV below the maximum of the XAS peak (E$_{in}=708$ eV, red arrow in Fig.~\ref{fig:fig1}). This is the incident energy used in the high-resolution RIXS spectrum of Fig.~\ref{fig:fig1}(d).

\begin{figure}
\centering
\includegraphics[scale=0.79,clip]{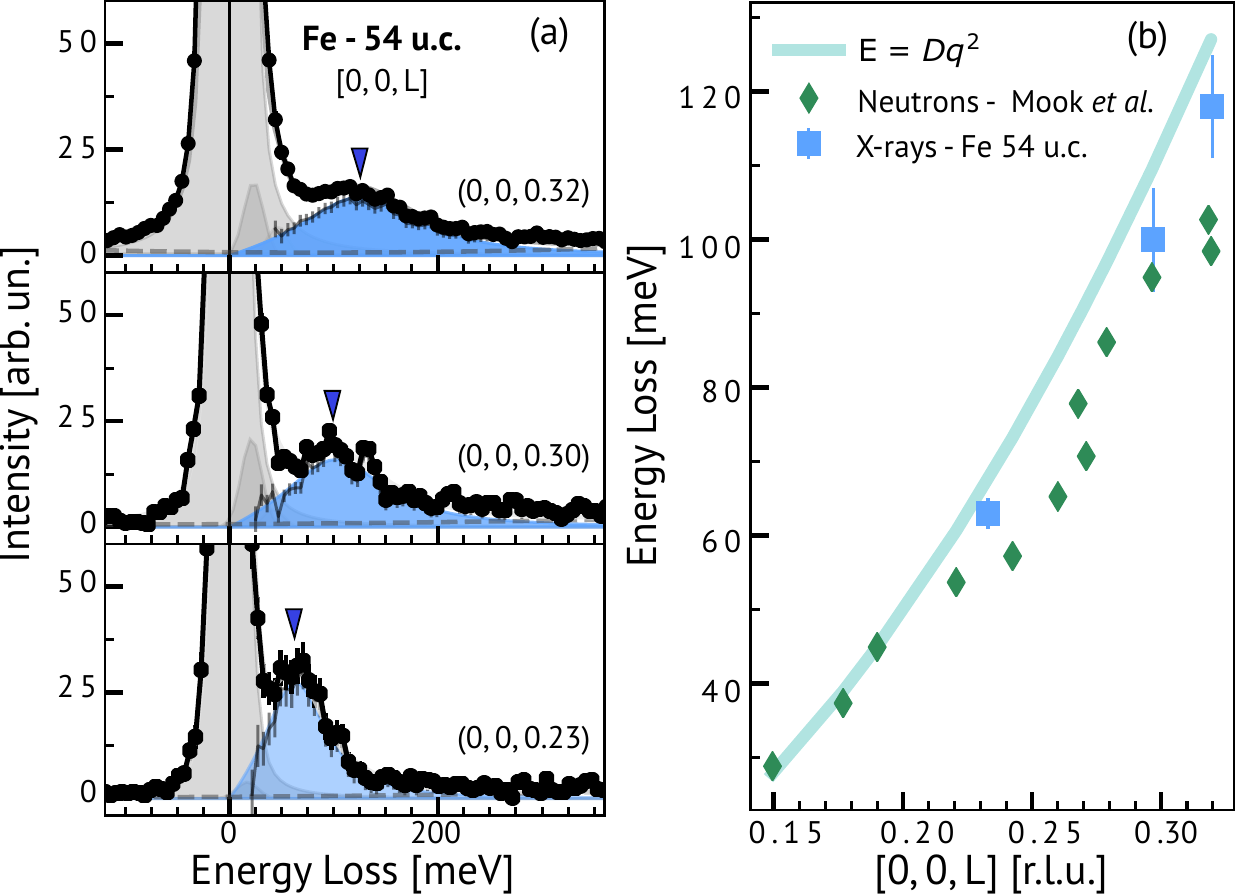}
\caption{\label{fig:fig2} \textbf{Momentum dependence of the ferromagnons in the bulk-like Fe film.} (a) RIXS spectra of a Fe 54~u.c. film along the [0,0,L] direction, for L = 0.23, 0.3 and 0.32 (black dotted line). The fitted contributions represented by filled areas are: the elastic line and the phonon in grey and the spin excitation in blue. Refer to the Supplementary Note 2 for details on the fitting components and the error bar definition. (b) Comparison of the dispersion of Fe measured by INS and RIXS along [0,0,L]. INS data are from Ref.~\onlinecite{mook_spinwave_1969} and are shown as green diamonds, RIXS data are blue squares with error bars. The error bar in the peak position is estimated in terms of a high-boundary corresponding to $0.5\cdot\gamma$, the width of the specular Lorentzian (see Supplementary Note 2). The aquamarine solid line represents the quadratic model  E = Dq$^2$ with D = 260 meV\AA$^2$ as proposed by Ref.~[\onlinecite{mook_spinwave_1969}]}
\end{figure}

To assess the origin and dispersion of the low-energy Raman-like peak, a momentum dependence investigation was performed by exploring three different $\bf q$ points along the [0,0,L] direction (with L = 0.23,0.30, and 0.32 r.l.u.). The results are reported in Fig.~\ref{fig:fig2}(a). The peak clearly disperses to higher energies for increasing $\bf q$. To quantify the energy dispersion of this mode, we fit the RIXS spectrum up to 500 meV using: $(i)$ a pseudo-Voigt function for the elastic peak at 0 meV; $(ii)$ a Gaussian function for the phonon around $17-25$ meV \cite{bergsma_normal_1967,minkiewicz_phonon_1967}); $(iii)$ a specular Lorentzian curve to fit the spectral weight from 30 to 300 meV; and $(iv)$ a quadratic background to account for the tail of the resonant fluorescence (details are reported in the Supplementary Note 2). The fitted contributions are represented as shaded filled curves in Fig.~\ref{fig:fig2}(a) for the 54~u.c. thick sample. The $\bf q$ dependence of the maximum of the specular Lorentzian is summarized in Fig.~\ref{fig:fig2}(b) as blue squares: this dispersion is in agreement with the spin dispersion of bulk $\alpha$-Fe measured by INS (green diamonds - raw data) and obeys a quadratic dispersion (aquamarine solid line) \cite{mook_spinwave_1969,collins_critical_1969,mook_neutron_1973,perring_highenergy_1991,paul_observation_1988}. Additionally, the broadening of the spin excitation profile observed by RIXS at large $\bf q$ is consistent with INS data, where it has been attributed to the coupling of spin excitations and Stoner continuum \cite{perring_highenergy_1991,paul_observation_1988}.

\begin{figure*}[t]
\includegraphics[width=0.97\textwidth, clip]{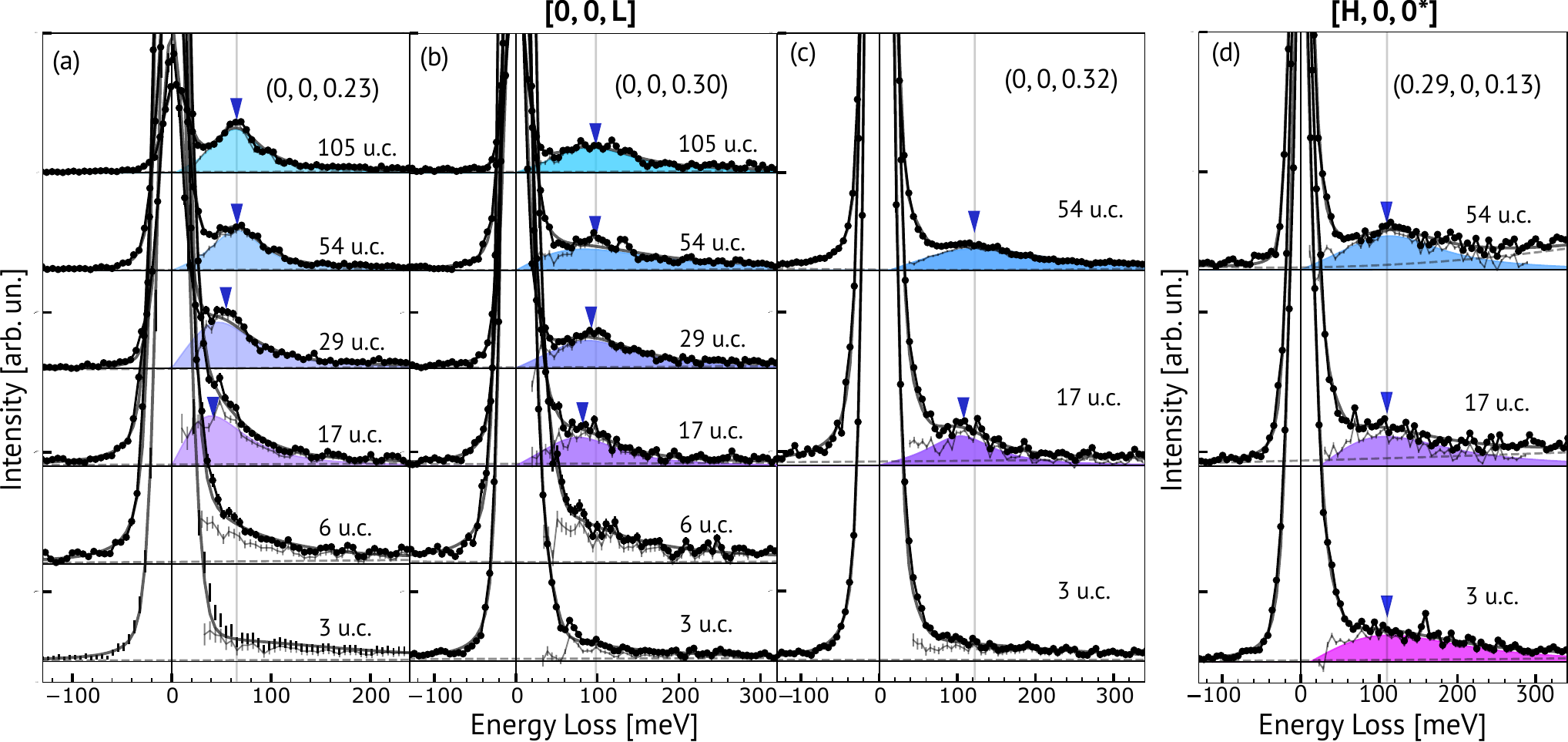}
\caption{\label{fig:fig3} \textbf{Anisotropic momentum dependence of the spin excitations in Fe films as a function of thickness.} (a-c) Out-of-plane [0,0,L] momentum dependence of RIXS spectra for  105~u.c., 54~u.c., 29~u.c., 17~u.c., 6~u.c. and 3~u.c. film thicknesses at L = 0.23, 0.3 and 0.32. (d) RIXS spectra at [H,0,0.13] in-plane momentum for a selection of thicknesses 54~u.c., 17~u.c. and 3~u.c.. In all plots we indicate the RIXS data as black dotted line, the fitted spectrum as thick solid grey line, the fitted specular Lorentzian with a filled curve, the residual (see Supplementary Note 2) with a thin solid grey line, and the fitted spin excitation peak position with a blue arrowhead.}
\end{figure*}

To characterize the dispersion of the spin excitations in three-dimensions, we recorded measurements for $\bf q$ mainly along the [H,0,0] in-plane direction. Due to kinematical constraints, $\bf q$ cannot be transferred solely along [H,0,0] but a small L component will also be present (L$\approx$0.13). In Figs.~\ref{fig:fig3}(a,c) and Supplementary Fig.~5 we exhibit the spectra at (0,0,0.3) and (0.29,0,0.13) which have degenerate energy and comparable line-shape. This degeneracy further corroborates the assignment of the low energy peaks to genuine $\alpha$-Fe spin excitations and demonstrates the three dimensional isotropy of the spin excitations in the 54 u.c. thick film. 

We now extend our presentation to thinner films, using the same experimental conditions of the 54 u.c. film. The results are summarized in Fig.~\ref{fig:fig3} for all the samples. We observe spin excitations down to the thinnest film. It is now important to separately discuss spectra for in-plane and out-of-plane momentum transfer. Along the out-of-plane direction [0,0,L] (panels (a-c) in Fig.~\ref{fig:fig3}), a clear peak in the $\sim 50-120$ meV range is present at all thicknesses down to 17 u.c.. Remarkably, we observe a gradual down renormalisation of the spin excitations while decreasing the thickness. The spin excitation energy is the same for 105 u.c. and the 54 u.c. films (with bulk-like dispersion). For 29 u.c. and 17 u.c., however, the spin excitation peak deviates from bulk-like behaviour to lower energies with an overall $\sim25\%$ reduction at $\bf q$ = (0,0,0.23) and (0,0,0.3) r.l.u. and a $\sim15\%$ reduction at $\bf q$= (0,0,0.32) r.l.u.. At $\bf q$ = (0,0,0.23), where the spin excitation peak is well defined, we notice  a slight decrease of the peak-width for films of thicknesses between 105 u.c. and 17 u.c. indicating a lower damping (we limit our analysis to film thicker than 17 u.c., see Supplementary Fig.~7 for details). At these thicknesses, a mesoscopic behaviour clearly emerges as a renormalisation of the magnon dispersion.

\begin{figure}[h!]
\centering
\includegraphics[scale=0.8,clip]{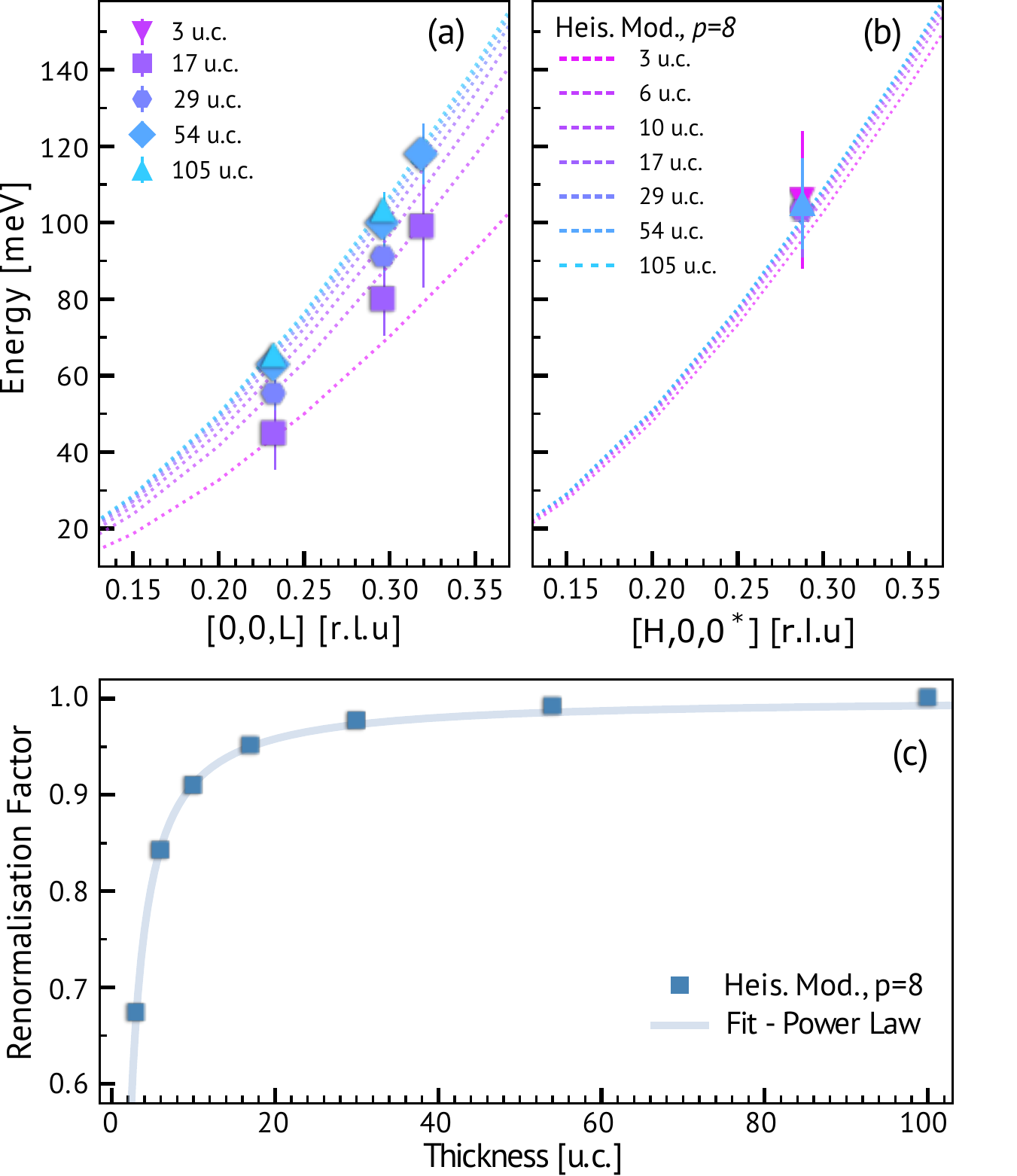}
\caption{\label{fig:fig4} \textbf{Confinement effect on the spin excitations in mesoscale Fe films.} (a,b) Summary of the extracted spin dispersion for different Fe film thicknesses along [0,0,L] (square-symbols) (a), and along [H,0,0$^*$] (round-symbols) (b). The error bars are defined as in Fig.~\ref{fig:fig2}(a).  Dashed lines are the dispersion curves calculated for different thicknesses using a Heisenberg model with short range interaction  $J_{ij}$ (p$=8$). (c) The $\bf q$-independent spin renormalisation factor versus thickness as extracted from our model along the [0,0,L] direction, and fitted with a power law function $1-A/x^k$, with $A= 8.22$ and $k=1.24$.}
\end{figure}

For the 6 and 3 u.c. films, the magnon dispersion in the out-of-plane direction cannot be similarly estimated, as the peak is no longer well defined and we find a $\sim2-3$ times larger peak-width. This broader peak-width seems unrelated to spin excitations as the spin excitations might be hidden underneath the elastic line (below 20-30 meV) and could be originated by residual particle-hole excitations dominating this energy range in absence of spin excitations. Because of this uncertainty, we do not further discuss the out-of-plane spin excitations for 6 and 3 u.c. out of the summary presented in Fig.~\ref{fig:fig4}(a).

The in-plane RIXS spectra for the 54 u.c., 17 u.c., and 3 u.c. films at (0.29,0,0.0$^*$) are reported in Fig.~\ref{fig:fig3}(d). No renoramalisation is observed for this direction, instead a peak with similar energy position, width, and intensity is present for all film thicknesses (down to 3 u.c.). The peak position around $\sim100$ meV matches the bulk-like dispersion of the spin excitations in $\alpha$-Fe. Therefore, this result shows that bulk-like in-plane spin excitations persist even in a few layers of Fe.

To interpret this result, we point out that the renormalisation of the spin excitations in the out-of-plane direction cannot be ascribed to strain. In the 3 u.c. film the strain along the out-of-plane direction is smaller ($\approx2\%$) than along the in-plane direction ($\approx3.8\%$), for which the measured spin excitations are unaffected by thickness. Thus, this is not a strain effect. We can also rule out effects such as substrate hybridization, and changes in magnetic moment. The hybridization of the Fe bands with the MgO substrate are unlikely to explain the spin excitations renormalisation observed in the relatively thick 29 u.c. film. Moreover, MgO is an insulator and lacks bands close to the Fermi level where the hybridization should affect the spin properties, making this scenario improbable. Moreover, our SQUID measurements of the Fe saturation magnetic moment confirm it is independent of the thickness, hence eliminating this possibility. Finally, we can disregard surface effects because the RIXS penetration depth is larger than the thickness of the film, implying that the signal stems from whole film. Therefore we can conclude that the scaling of the spin excitations has to be interpreted as a confinement effect.

To support this mechanism, we consider a long-range, isotropic classical Heisenberg Hamiltonian on a cubic lattice. The effective exchange interaction between sites $i$ and $j$ in this model Hamiltonian (see Supplementary Note 3 for more details) is ~
\begin{equation}\label{eq:eq1}
J_{ij} = J_{0} \left ( \frac{a}{R_{ij}} \right )^p \, ,
\end{equation}

\noindent where $R_{ij}$ is the distance between the two sites, $a$ is the lattice constant, $J_{0}$ sets the strength of the exchange interaction and $p$ determines the effective length scale of the interaction. This formulation was previously used in theoretical works to describe the spin interactions in Fe model systems \cite{magnus_long-range_2016,pajda_ab_2001}, effectively modeling the itinerant electron behaviour.

Figures \ref{fig:fig4}(a) and (b) show as solid lines the out-of-plane and in-plane spin excitations dispersion obtained from evaluating the model with a short-range $J_{ij}$ interaction ($p=8$, see Ref. [\onlinecite{turek_exchange_2006}] and Supplementary Note 3.). Interestingly, this simple, isotropic exchange model is able to qualitatively reproduce the anisotropic thickness-dependent dispersion observed in our experiments (square and round markers in Fig. \ref{fig:fig4}(a,b)). In particular, the out-of-plane renormalisation is due to the reduction of the near-neighbor spins at the film surfaces. The isotropic exchange interaction together with the loss of Fe magnetic interactions at the Fe film surfaces produces an apparent anisotropic reduction of $J$: The calculations reveal that this effect impacts the energies of the out-of-plane modes far more than the in-plane ones. In contrast, when using a long-range $J_{ij}$ interaction ($p=4$), we find that the in-plane modes scale down similarly to the out-of-plane ones (see Supplementary Note 3 and Supplementary Fig. 10.), in contradiction with the experimental trends.

While this simple model correctly predicts the qualitative behaviour, it  misses a complete one-to-one correspondence with the data as a function of thickness, the outcome of this theory is twofold: First, it proves that the confinement effect, associated with the geometrical loss of Fe spins due to the limited thickness, causes a renormalisation of the spin excitations. Second, it supports the short-range nature of the magnetic interactions in Fe, responsible for the anisotropic spin excitation evolution between out-of-plane and in-plane directions. Our direct measurement of the length scale of spin interactions $J_{ij}$ is consistent with INS experiments which fitted the dispersion using nearest-neighbor Heisenberg model \cite{shirane_measurement_1965,mook_neutron_1973}.

These results demonstrate that the confinement effect can be used as a knob to tune the spin excitations in meso-scaled systems. This tuning can be parameterized by noting that the calculated dispersion curves for each thickness in Fig.~\ref{fig:fig4}(a), can be scaled by a $\bf q$-independent factor, which we call the renormalisation factor in Fig.~\ref{fig:fig4}(c). The renormalisation factor follows a power law behaviour as a function of thickness. By knowing the exchange coupling of the bulk material, this factor can be used to predict the thickness dependence of the spin excitations in cubic systems. Being based only on geometrical arguments, the scaled dispersion can be extrapolated to low energy spin excitations (in the order of GHz) which are more commonly employed in magnonics, extending the applicability range of our investigation. 
More generally, in applications where magnetic mediums are used to propagate the signal, our study demonstrates the use of confinement to select the appropriate energy of spin-waves to be employed for specific purposes. Our approach is analogous to the modification of refractive index in optics, where the film thickness acts to tune the `magnetic' refractive index of the medium in magnonics.

In summary, we combine MBE synthesis and RIXS experiments to unveil the thickness dependence of the spin excitations in Fe films covering almost two orders of magnitude in thickness. In bulk-like films, RIXS is used for the first time to measure ferromagnetic spin excitations consistently with INS. From this demonstration, we extend our study to mesoscopic thin films and uncover a remarkable renormalisation of the out-of-plane spin excitations as the film thickness is decreased down to few unit cells. In contrast, in-plane spin excitations are thickness independent preserving a bulk dispersion down to 3 u.c.. These trends can be explained using a Heisenberg-like formalism based on a short-range exchange isotropic interaction and accounting for confinement through the loss of Fe spin interactions at the film surface. Using this theory we are able to predict the dependence of the spin excitations as a function of thickness. Our predicitons can be applied in other systems to realize tunable spin excitations for which confinement is the dominant effect. From an application perspective, our study shows a method to tailor the spin excitation energies with impact on spintronics and magnonics by providing an effective way to tune the refractive index in magnetic materials, equivalent to manipulating the dielectric function in photonics.

\section{Methods}
\textit{Sample preparation}
High-quality, crystalline films of Fe with varying thickness were prepared by molecular beam epitaxy (MBE). The Fe [001](110)-oriented films were grown epitaxially on MgO[001](100) substrates with a 45\textdegree~ rotation around the [001] axis at a substrate temperature of 300 \textdegree{C}. During the growth, the films were monitored using \textit{in-situ} reflection high-energy electron diffraction (RHEED) to assess the crystalline quality (see Supp. Info. for details on the growth and characterization of the films). A $\sim5$~nm protective layer of MgO was deposited to avoid oxidation. 
More details of the sample growth and characterization are reported in the Supplementary Note 1.

\textit{XAS and RIXS measurements}
X-ray Absorption Spectroscopy (XAS) was measured at the SIX 2-ID beamline of the National Synchrotron Radiation Facility II (NSLS-II) with a resolution of $\approx50$ meV. XAS at the Fe $L_3$ edge ($\sim707$~eV) was measured in partial fluorescence yield (PFY) integrated over an energy window of 670-720 eV and acquired using the RIXS spectrometer.

High-resolution RIXS experiments were performed at the SIX 2-ID beamline of NSLS-II, using the Centurion RIXS spectrometer \cite{dvorak_towards_2016}. The combined energy resolution at the Fe-L$_3$ edge ($\approx708$ eV) was $\Delta E = 23$~meV (FWHM) throughout the experiment. $\pi$-polarized light was used to reduce the elastic scattering signal. 

All XAS and RIXS data were measured at 100~K.

For the detailed analysis of the RIXS data see Supplementary Note 2.

\textit{Calculations}
Details of the calculations are reported in the Supplementary Note 3. 

\section{Authors contribution}
 V.~B. conceived the project. J.~P., S.~L., J.~L., Y.~G., A.~B., I.~J., and V.B performed the RIXS experiments. S.~J., C.~H.~A, and F.~J.~W. prepared the Fe/MgO thin films and characterized them using SQUID and x-ray reflectometry. V.~B. analysed and interpreted the data, with the help of J.~P.. K.~G. performed the theory calculations. J.~P. and V.~B. wrote the manuscript with input from all the authors.

\section{Acknowledgements}
The authors are indebted to John Hill, Claudio Mazzoli, John Tranquada, Tim Ziman, Mark Stiles, Patrick Bruno, and Elio Vescovo for fruitful discussions.
This work was supported by the U.S. Department of Energy, Office of Science, Basic Energy Sciences, Early Career Award Program.  Work at Yale University was supported by the U.S. Department of Energy, Office of Science, Office of Basic Energy Sciences under Award No. DE-SC0019211. K.G. was supported by the US Department of Energy, Office of Science, Basic Energy Sciences as part of the Computational Materials Science Program. This research used beamline 2-ID of the National Synchrotron Light Source II, a U.S. Department of Energy (DOE) Office of Science User Facility operated for the DOE Office of Science by Brookhaven National Laboratory under Contract No. DE-SC0012704. SXRD measurements were performed at the beamline 33-ID-D of the Advanced Photon Source, a U.S. Department of Energy (DOE) Office of Science User Facility operated for the DOE Office of Science by Argonne National Laboratory under Contract No. DE-AC02-06CH11357.

 \section{Correspondence} 
 Correspondence and requests for materials should be addressed to J. Pelliciari and V. Bisogni.
 
\section{Competing Interests} 
The authors declare no competing and financial interests.

\section{Data Availability}
Data that support the findings of this study are available upon reasonable request from the corresponding authors.

\end{document}